\documentclass[a4paper, twoside, 12pt]{article}
\usepackage[utf8]{inputenc}
\usepackage[T1]{fontenc}
\usepackage{graphicx}
\usepackage{tabularx,longtable}
\usepackage{hyperref}
\usepackage{caption}
\usepackage{url}

\usepackage[left=2.5cm, right=2.5cm, top=2.5cm, bottom=2.5cm, bindingoffset=1.5cm, head=15pt]{geometry} 
\usepackage{setspace}
\usepackage{fancyhdr,amsmath,amssymb}
\usepackage{algpseudocode}
\usepackage{algorithm}
\usepackage{cleveref}

\title{Analysis of Linux-PRNG (Pseudo Random Number Generator)}

\newcommand{\thesisauthor}{Ayush Bansal} 
\newcommand{\studentID}{160177} 
\newcommand{\thesistype}{UGP Report} 
\newcommand{\supervisor}{Dr. Pramod Subramanyan \& \newline Dr. Satyadev Nandakumar}

\newcommand{\toolname}{\textsc{Entropy-Mixer}}

\begin{document}

\makeatletter
\begin{titlepage}
    \begin{center}
        \vspace*{1cm}

        \Large
        \textbf{\@title}

        \vspace{1.5cm}
        
        \thesistype{}
        
        \vspace{1cm}

        \begin{figure}[htbp]
             \centering
             \includegraphics[width=.5\linewidth]{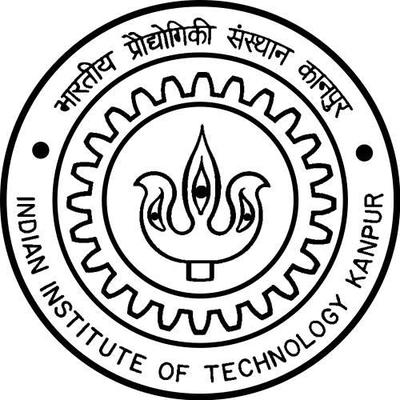}
        \end{figure}

        \vspace{1cm}

        \large
        \textbf{Author}: \thesisauthor{} (Student ID: \studentID{})\\
        \large
        \textbf{Advisor}: \supervisor{}\\

        \vspace{1cm}
        \large
        Department of Computer Science and Engineering\\
        Indian Institute of Technology, Kanpur\\

        \vspace{1cm}
        June 16, 2020

    \end{center}
\end{titlepage}
\makeatother


\clearpage
\thispagestyle{empty}
\section*{Abstract}

The Linux pseudorandom number generator (PRNG) is a PRNG with entropy inputs and is widely used in many security-related applications and protocols. This PRNG is written as an open-source code which is subject to regular changes. It has been analysed in the works of Gutterman \textit{et al.} \cite{gutterman}, \textit{Lacharme et al.} \cite{prng-revisited} and \cite{prng-perf}, while in the meantime, several changes have been applied to the code, to counter the attacks presented since then. Our work describes the Linux PRNG of kernel versions 5.3 and upwards. We discuss the PRNG architecture briefly and in detail about the entropy mixing function.

Our goal is to study the entropy mixing function and analyse it over two properties, namely, injectivity and length of the longest chain. For this purpose, we will be using SAT solving and model counting over targetted formulas involving multiple states of the Linux entropy store.

\clearpage
\pagenumbering{Roman}
\tableofcontents
\clearpage

\pagenumbering{arabic}


\clearpage

\section{Introduction}
\label{sec:Intro}

Many security protocols base their security on the impossibility of an attacker being able to guess the random data, such as session keys for cryptosystems or nonces for cryptographic protocols. However, the amount of required random data can differ significantly with the application. Therefore, random data generation should take into account the fact that the user can request either high-quality random data or a great amount of pseudorandom data. Even a recent article \cite{lwn-article} demanding a fix in Linux's \texttt{getrandom()} function to provide an alternative to get a great amount of pseudorandom data without getting blocked for entropy.

A pseudorandom number generator with entropy inputs produces bits non-deterministically as the internal state is frequently refreshed with unpredictable data from one or several external entropy sources. It is typically made of (1) several physical sources of randomness called entropy sources, (2) a harvesting mechanism to accumulate the entropy from these sources into the internal state, and (3) a post-processing procedure. The latter frequently uses a \textit{cryptographically secure PRNG} (CSPRNG) to generate outputs and to update the internal state.

\begin{figure}[h]
    \includegraphics[scale=0.4]{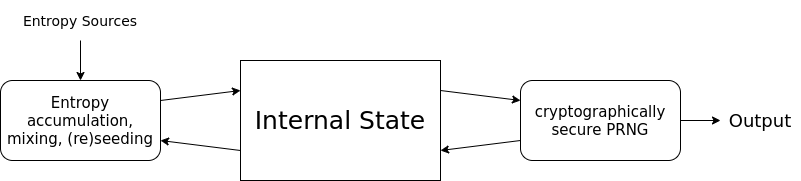}
    \centering
    \caption{PRNG Model}
\end{figure}

In Unix-like operating systems, a pseudo-random number generator with input was implemented for the first time for Linux $1.3.30$ in $1994$. The entropy source comes from device drivers and other sources such as latencies between user-triggered events (keystroke, disk I/O, mouse clicks, ...). It is gathered into an internal state called the \textit{entropy pool}. The internal state keeps an estimate of the number of bits of entropy in the internal state and (pseudo-)random bits are created from the special files \texttt{/dev/random} and \texttt{/dev/urandom}. Barak and Halevi \cite{barak} discussed the generator \texttt{/dev/random} briefly, but its conformity with their robustness security definition is not formally analysed.

The first security analysis of these generators was given in $2006$ by Gutterman, Pinkas and Reinman \cite{gutterman}. It was completed in $2012$ by Lacharme, Röck, Strubel and Videau \cite{prng-revisited}. Gutterman \textit{et al.} \cite{gutterman} presented an attack based on kernel version $2.6.10$ for which a fix was published in the following versions. Lacharme \textit{et al.} \cite{prng-revisited} gives a detailed description of the operations of the generators and provides a result on the entropy preservation property of the \textit{mixing function} used to refresh the internal state.

In the next few sections, we will discuss the \textit{Architecture} of the Linux PRNG briefly and in detail the \textit{mixing function} applied over the \textit{entropy pool} to refresh the internal state of the random number generator. The works of Lacharme \textit{et al.} \cite{prng-revisited} suggested that the \textit{mixing function} has the property of entropy preservation, and analysing the same property is one of our goals. \newline

\noindent The motivations behind this analysis are:
\begin{itemize}
  \setlength\itemsep{0em}
  \item Gain insights into the working of the Linux PRNG
  \item Gain a deeper undestanding of the entropy estimation and accumulation processes.
  \item Verify the entropy preservation properties as claimed in \cite{prng-revisited}.
\end{itemize}

\noindent Previous works on the analysis of Linux PRNG claim that the \textit{mixing function} which is used to refresh the internal state preserves the entropy received from various entropy sources, our primary goal with this analysis is to verify this using the techniques of SAT solving and model counting. \newline

\noindent Our goals regarding this analysis are:
\begin{itemize}
  \setlength\itemsep{0em}
  \item Comment the property of injectivity of the entropy mixing function.
  \item Estimate the length of the longest chain, i.e. the size of input bytes which when supplied with an initial state of \textit{entropy pool} to the mixing function leaves the entropy pool unchanged. 
\end{itemize}

\clearpage
\section{Architecture}
\label{sec:Architecture}

The Linux PRNG has been a part of the Linux kernel since $1994$. Ts'o wrote the original version, and Mackall \cite{mackall} later modified it. The generator, apart from the entropy input hooks inserted into drivers, represents about $2500$ lines of C code in a single source file, \texttt{drivers/char/random.c}.

\subsection{General Structure}

\begin{figure}[h]
    \includegraphics[scale=0.35]{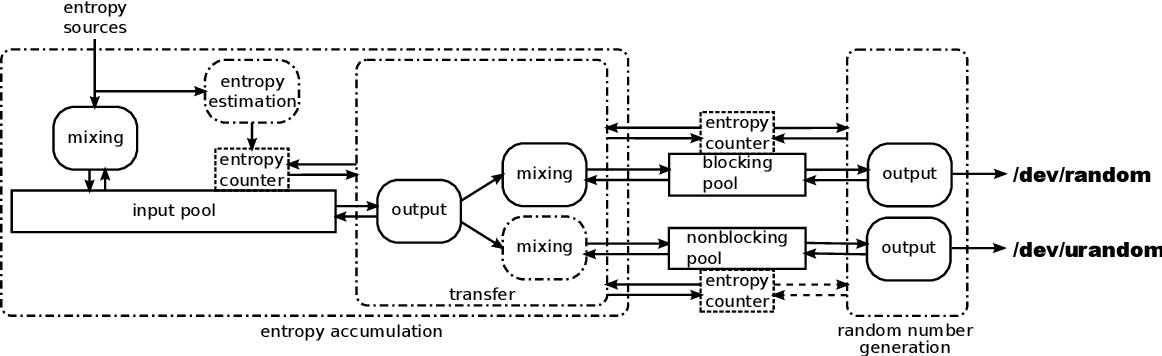}
    \centering
    \caption{PRNG Model \cite{prng-revisited}}
    \label{fig:prng}
\end{figure}

Unlike other PRNGs, the internal state of the Linux PRNG is divided into three entropy pools, the \textit{input pool}, the \textit{blocking pool} and the \textit{non-blocking pool}. The blocking and the non-blocking pools together form the \textit{output pools}. The size of the input pool is $128$ $32$-bit words ($4096$ bits) and the size of each output pool is $32$ $32$-bit words ($1024$ bits). The generator is designed to perform the collection of entropy inputs as efficiently as possible; therefore, it uses a linear mixing function instead of a more usual hash function. The general structure of a Linux PRNG is shown in \Cref{fig:prng}.

The PRNG relies on external entropy sources. The various system events like keyboard and mouse movements, disk events, and hardware interrupt inside the kernel act as the entropy source for Linux PRNG and inputs are collected from them. The collected inputs are accumulated inside the input pool. An entropy counter is associated with each pool, decremented when entropy is extracted from the pools and incremented when entropy is added to the pools. Entropy transfers take place from the input pool to the output pools only when some entropy needs to be extracted from Linux PRNG. Hence, the entropy counters of the output pools remain close to zero except for the transition state. It makes the entropy counter of the input pool as the most significant one as the available entropy in Linux PRNG is reflected by the entropy counter value of the input pool \cite{prng-revisited}.

Linux PRNG provides two character interfaces \texttt{/dev/random} and \texttt{/dev/urandom} to read the random output. The \texttt{/dev/random} reads from the blocking pool and \texttt{/dev/urandom} read from the non-blocking pool. \texttt{/dev/random} provides cryptographically stronger bytes, blocks if sufficient entropy is not present in the pools and resumes when enough new entropy samples have been mixed into the input pool. Thus, it limits the number of generated bits according to the estimation of the entropy available in the PRNG. \texttt{/dev/urandom} always provides the desired amount of bytes and does not block; it is meant for the fast generation of large amounts of random data. In addition to these two character interfaces, Linux PRNG also provides \texttt{get\_random\_bytes()} which is a kernel interface. It allows the other kernel components to read from the non-blocking pool.

Entropy accumulation in Linux PRNG is a process of three major steps, namely, collecting entropy from various entropy sources, mixing the collected entropy into the input pool using a \textit{mixing function} and estimating the total entropy associated with each input which is mixed into the input pool. Broadly, the structure of Linux PRNG consists of four major components which are described in the following subsections. \Cref{ent-input} describes the various entropy sources briefly used by the Linux PRNG for accumulating entropy. \Cref{sec:mixing} describes the mixing function used for mixing the inputs collected from different entropy sources into the input pool. The same mixing function is also used when entropy is transferred from the input pool to the output pools when the entropy content of the output pools is less than that of the requested entropy bytes from Linux PRNG. \Cref{ent-est} describes briefly the entropy estimation process used by Linux PRNG to estimate the entropy associated with each input which is mixed into the input pool. \Cref{ent-out} briefly describes the output generation process in Linux PRNG, which is used whenever entropy is extracted from the output pools or a transfer of entropy occurs between the input pool and the output pools.

\subsection{Entropy Sources} \label{ent-input}

Entropy inputs are injected into the generator for initialization, and through the updating mechanism, provides the backbone of the security of the PRNG. The Linux PRNG is intended to be used independently of any specific hardware. Therefore, it cannot rely on nondeterministic physical phenomena generally used in the random generation, which require additional hardware.

\begin{figure}[H]
    \includegraphics[scale=0.5]{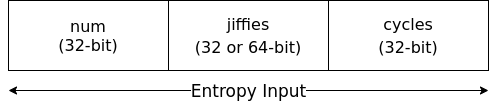}
    \centering
    \caption{An Entropy Input}
    \label{fig:ent-input}
\end{figure}

The Linux PRNG processes events from different entropy sources, namely, user inputs (keyboard and mouse movements), disk events, and hardware interrupts. Each input generated from these entropy sources are mixed into the input pool and consists of three values, as shown in \Cref{fig:ent-input}, also known as \textit{sample}.

In the above \textit{sample},
\begin{itemize}
  \setlength\itemsep{0em}
  \item \textbf{num}: It is a $32$-bit value representing the specific event type.
  \item \textbf{jiffies}: It represents the jiffy count value of the system at which the events mixed into the input pool of the Linux PRNG. `Jiffy' is the time between two ticks of the system timer interrupt, and the jiffy count value of a system represents the total number of ticks of system timer interrupt from the boot time. 
  \item \textbf{cycles}: It is a $32$-bit value representing the CPU cycle count value.
\end{itemize}

The functions \texttt{add\_input\_randomness}, \texttt{add\_disk\_randomness} and \texttt{add\_interrupt\_randomness} of \textit{/drivers/char/random.c} file of Linux kernel manage the collection of inputs from different entropy sources of the Linux PRNG. All these inputs are accumulated into the input pool of Linux PRNG through a mixing function.

\subsection{Entropy Estimation} \label{ent-est}

The entropy estimation is one of the crucial steps for generation of data via \texttt{/dev/random}. An accurate estimation of whether the corresponding pool contains enough entropy to generate unpredictable output data is essential.

The entropy counter associated with each pool keeps an estimate of the entropy content of that pool. Entropy estimation of a Linux PRNG is based on the timing of the events generated from different entropy sources, which mixed into the input pool. Although, both \textit{cycles} and \textit{jiffies} represents the timing of the event, \textit{jiffies} being the one having more coarser granularity is used for estimating the entropy of Linux PRNG. The difference of \textit{jiffies} associated with each input event is used for entropy estimation.

\subsection{Output Generation} \label{ent-out}

There are three output interfaces provided by the Linux PRNG for extracting entropy from the blocking and non-blocking pools. \texttt{/dev/random} and \texttt{/dev/urandom} are the character interfaces for the users and \texttt{get\_random\_bytes()} is the kernel interface for extracting entropy. \texttt{/dev/random} extracts entropy from blocking pool and may block if enough entropy is not present, providing very high-quality random bits. \texttt{/dev/urandom} can provide as many bits as requested, extracting from the non-blocking pool. The function \texttt{get\_random\_bytes} is available only within the kernel and extracts entropy from the non-blocking pool.

The output generation function can be divided into two phases, namely, feedback phase and extraction phase. It uses \texttt{SHA-1} hash function in two-steps.

\begin{enumerate}
  \setlength\itemsep{0em}
  \item All bytes of the pool are hashed to produce $5$-word hash and $20$ bytes are mixed back into the pool using mixing function. Mixing back of generated hash in the pool prevents the backtracing attacks. This step forms the feedback phase of output generation function.
  \item $16$ words from the mixed pool content are again hashed along with the 5 output words from the previous step. The 20-byte output from this second hash is folded in half to get a $10$ byte output.
\end{enumerate}

The entropy counter of the output pool is finally decremented by the number of bytes generated from the Linux PRNG.

\clearpage
\section{Mixing Function}
\label{sec:mixing}

\begin{figure}[h]
    \includegraphics[scale=0.4]{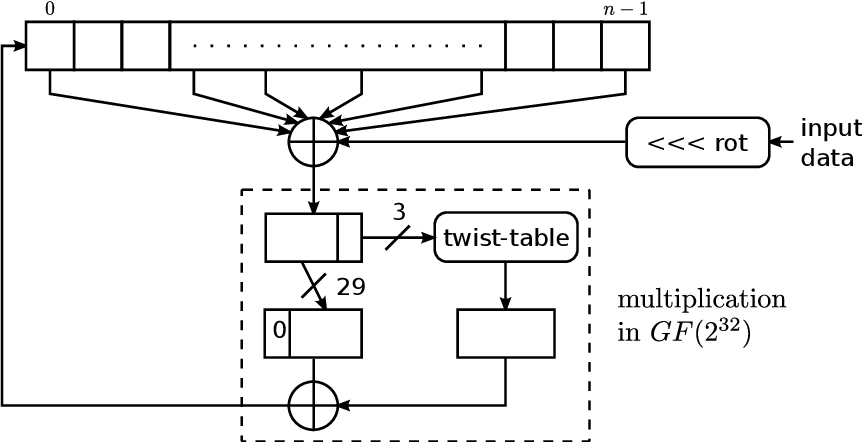}
    \centering
    \caption{The mixing function \cite{prng-revisited}}
    \label{fig:mixing}
\end{figure}

The structure of the mixing function of Linux PRNG is shown in \Cref{fig:mixing}. This function mixes one byte of the input at a time by first extending that one byte to a $32$-bit word, followed by a rotation of the extended $32$-bit word by a changing factor, and finally mixing it in the pool by using a linear shift register. It is designed so that no entropy gets lost while diffusing into the pools.

The Mixing function is defined by the function \texttt{mix\_pool\_bytes} in \newline
\noindent \textit{/drivers/char/random.c} file of Linux kernel. The function \texttt{mix\_pool\_bytes} takes three arguments.

\begin{enumerate}
  \setlength\itemsep{0em}
  \item \textbf{struct entropy\_store *r}: Entropy store is a structure which contains various entities such as \texttt{poolinfo} containing the definition for both input and output pools, \texttt{pool} containing the name of the pool being used for mixing the entropy input, \texttt{entropy count} representing the entropy count of the respective pool. The input pool is declared as an array containing $128$ words and the output pools which includes both blocking and non-blocking pools are declared arrays of $32$ words each. The variables \texttt{add\_ptr} and \texttt{input\_rotate} are defined as data of type unsigned int (C data type).
  \item \textbf{const void *in}: Pointer to the input collected from various entropy sources which is mixed into the entropy pools. When inputs are mixed into the input pool from the function \texttt{add\_timer\_randomness}, \texttt{in} becomes a pointer to the structure \texttt{sample} as shown in \Cref{fig:ent-input}.
  \item \textbf{int nbytes}: This represents the total number of bytes to be mixed into the pool. The total number of bytes added to the input pool is either $12$ bytes or $16$ bytes depending on the architecture of the system used.
\end{enumerate}

The mixing function updates the input entropy store after mixing the input bytes received from various entropy sources.

Apart from these function arguments, there are also different variables declared inside the mixing function. \texttt{unsigned long integers tap1, tap2, tap3, tap4} and \texttt{tap5} are declared inside the function, their values are $104,76,51,25$ and $1$ respectively for the input pool and $26,19,14,7$ and $1$ respectively for the output pools. A variable \texttt{wordmask} is also defined inside the function \texttt{mix\_pool\_bytes} whose value is $127$ for the input pool and $31$ for the output pools.

\subsection{The Process}

The input bytes also known as samples are mixed into the input pool using mixing function, one byte at a time. For a $n$ byte sample, the whole process of mixing is repeated $n$ times. The one-byte value of the sample is first type casted to \texttt{\_\_u32} (expanded to $32$ bits) form of C language and is then rotated with a function called \texttt{rol32}, defined in \textit{linux/bitops.h} of Linux kernel. The \texttt{rol32} function takes two arguments, a \texttt{word} to be rotated ($32$ bit) and a factor \texttt{shift} by which that word is rotated.
$$\text{w} = (\text{word} \gg \text{shift})\ |\ \text{word} \ll (32-\text{shift})$$

The initial value of \texttt{shift} is $0$. The variable \texttt{i} declared inside the mixing function is equal to the output of a function \texttt{ACCESS\_ONCE(x)}, where \texttt{x} is equal to a variable \texttt{add\_ptr} defined in the structure \texttt{entropy store}, this function is defined as a macro in \textit{linux/compiler.h} of Linux kernel. In each iteration, a \texttt{byte} of the entropy input is mixed into the entropy pool via following steps:
\begin{enumerate}
  \setlength\itemsep{0em}
  \item The \texttt{byte} representing one byte value of the input sample is expanded to $32$ bits, rotated using the \texttt{rol32} function and the variable \texttt{w} is assigned its value.
  \item The variable \texttt{i}, which was assigned the value \texttt{add\_ptr} from the entropy store is updated with \texttt{(i - 1) \& wordmask}.
  \item The word \texttt{w} is then $XOR$ed at each of the tap locations defined for the corresponding pool in the following way:
    \begin{align*}
      w &= w \oplus r\rightarrow \text{pool}[i]; \\
      w &= w \oplus r\rightarrow \text{pool}[(i+\text{tap1})\ \&\ \text{wordmask}]; \\
      w &= w \oplus r\rightarrow \text{pool}[(i+\text{tap2})\ \&\ \text{wordmask}]; \\
      w &= w \oplus r\rightarrow \text{pool}[(i+\text{tap3})\ \&\ \text{wordmask}]; \\
      w &= w \oplus r\rightarrow \text{pool}[(i+\text{tap4})\ \&\ \text{wordmask}]; \\
      w &= w \oplus r\rightarrow \text{pool}[(i+\text{tap5})\ \&\ \text{wordmask}];
    \end{align*}
    The word \texttt{w} is then finally mixed back into the entropy pool at the $i^th$ location using the following:
    $$r\rightarrow \text{pool}[i] = (w \gg 3) \oplus \text{twist\_table}[w\ \&\ 7];$$

    Twist table is defined in \textit{drivers/char/random.c} as an array containing $8$ $32$-bit hexa-decimal values. These values are the monomials in the field $(F_2)/(Q)$ \cite{koopman}. $Q(x)$ is defined in \cite{koopman} as equal to $x^{32} + x^{26} + x^{23} + x^{22} + x^{16} + x^{12} + x^{11} + x^{10} + x^8 + x^7 + x^5 + x^4 + x^2 + x + 1$, is a CRC32 polynomial used for Ethernet protocol.
    
  \item Finally, the \texttt{input\_rotate} value, i.e. \texttt{shift} is increased by a value of $7$ if the value of \texttt{i} is non-zero, else it is incremented by 14.
\end{enumerate}

When all the bytes of an entropy input are mixed into the entropy pool, the value of \texttt{input\_rotate} and \texttt{i} is then again accessed using the function \texttt{ACCESS\_ONCE} and is assigned to the respective variables.

\clearpage
\section{Design and Evaluation}
\label{sec:analysis}

Our main goal, as discussed in \Cref{sec:Intro} to analyse the Linux's \textit{mixing function} and comment on the properties of injectivity and length of the longest chain. With the help of this analysis, we can verify whether the entropy once supplied to the \textit{mixing function} is preserved or not. 

For our analysis, we will use the techniques of SAT solving and model counting in order to ensure that the function preserves entropy. We have imitated the parts of code in \textit{/drivers/char/random.c} concerning the \textit{mixing function} and the \textit{entropy store} in \textit{scala} as part of our tool \toolname. In the next subsection, we are going to discuss in detail the implementation design of \toolname.

\subsection{Design}
Our tool \toolname, takes an initial entropy store and an array of entropy bytes as input and applies the mixing function on the store, mixing one byte at a time and updates the initial entropy store to the final state of the entropy store after the mixing is complete. During this entire process, the variables involved are \textit{bit-vector expressions}. Hence, the final store generated is also a bit-vector expression represented as an expression of input entropy store and entropy bytes combined with logical bit operations. \newline

\noindent The model structure of our tool is as follows:
\begin{enumerate}
  \setlength\itemsep{0em}
  \item \textbf{Expr Class}: Denotes a boolean expression containing boolean variables and boolean operators (each defined as a case class), an expression can be one of the following.
    \begin{itemize}
      \setlength\itemsep{0em}
      \item A constant boolean value, i.e. \textit{true} or \textit{false}.
      \item A boolean variable (value unknown), denoted by a name (string).
      \item Complement of an expression, i.e. \texttt{Not(expr)}.
      \item Boolean operator applied over a list of expressions, such as \textit{and}, \textit{or} and others.
    \end{itemize}
  \item \textbf{BitVector Class}: Denotes a bit-vector of arbitrary but constant length defined as an array of boolean expressions (using the \textit{Expr} class).
  \item \textbf{Entropy Store}: Denotes the \textit{entropy pool} as discussed in \Cref{sec:mixing}, the entropy pool can be of size $128$ $32$-bit words in which case it is an input pool, or it can be of size $32$ $32$-bit words becoming one of the output pools. Each word is denoted as an instance of the \textit{BitVector} class.
\end{enumerate}

\noindent The mixing function involves a good amount of bit operations to be performed on the data, in order to incorporate those in our model, we have designed \textit{BitVectorOperations} object to address our needs of bit operations. The operations defined are as follows:
\begin{enumerate}
  \setlength\itemsep{0em}
  \item \textbf{bitOr, bitAnd, bitXor} and \textbf{bitXnor}: These operations take $2$ bit-vectors of same size (e.g. $32$ bit) and generate a bit-vector after applying bitwise \textit{or, and, xor} and \textit{xnor} respectively on the $2$ bit-vectors.
  \item \textbf{bitLeftShift} and \textbf{bitRightShift}: These operations take a bit-vector and an integer value, applies the bitwise \textit{left shift} and \textit{right shift} respectively to generate the output bit-vector.
  \item \textbf{expandBitVector}: It takes as input a bit-vector and an integer value (greater than size of bit-vector), expands the bit-vector to the provided size.
  \item \textbf{threeMux}: It takes an array of bit-vectors and a bitvector \textit{select} value, and it imitates the functionality of a $3$ pin multiplexer.
\end{enumerate}

Using the above classes and operations, we imitate the code of \texttt{mix\_pool\_bytes} function of Linux kernel in order to gain the final \textit{entropy store} in the form of \textit{bit-vector expressions} to be used further by SAT solvers.

\subsection{Evaluation}

We have discussed in detail the design of our tool and model setup, in this subsection, we will look at the functions we tried to analyse using SAT solvers to judge the property of \textit{injectivity} for the mixing function.

We used \textit{Cadical} \cite{cadical} for SAT Solving and \textit{Maxcount} \cite{maxcount} solver for \textit{Maximum Model Counting} during our analysis. \newline

\noindent The formulas that we analysed are as follows:
\begin{enumerate}
  \setlength\itemsep{0em}
  \item \textbf{Formula-1}: \begin{equation} (i_1 \neq i_2)\ \& \ (f(x,i_1) = f(x,i_2)) \label{eq:1} \end{equation}
    This formula checks the condition of $2$ different sets of entropy input bytes (with the same size) being used as input for the mixing function (denoted by $f$ here) and still producing the same state of entropy store.
  \item \textbf{Formula-2}: \begin{equation} ((x_1 \neq x_2)\ |\ (i_1 \neq i_2))\ \&\ (f(x_1,i_1) = f(x_2,i_2)) \label{eq:2} \end{equation}
    This formula checks the condition of the final state of entropy stores being same after the mixing even though the initial state of entropy stores was different or the entropy bytes used for mixing were distinct (of same size but different value).
\end{enumerate}

For testing above formulas, we used the output pool as the entropy store, the size of the output pool is $32$ $32$-bit words, the input entropy bytes will be an array of size $8$ bit-vectors. \newline

\noindent After running SAT Solvers on the above formulas:
\begin{enumerate}
  \setlength\itemsep{0em}
  \item \textbf{Formula-1}: \textbf{UNSAT} upto size $88$ of input entropy bytes, i.e. variables $i_1$ and $i_2$.
  \item \textbf{Formula-2}: \textbf{SAT} upto size $288$ of input entropy bytes.
\end{enumerate}

\clearpage
\section{Conclusion/Future Work}
\label{sec:conclusion}

We presented our design for \toolname which takes as input an initial state of the \textit{entropy store} and an array of entropy bytes, applies the mixing function over the entropy store, one byte at a time and produces the final state of the entropy store. \toolname was used in analysing the property of \textit{injectivity} for the mixing function.

We targetted to comment on the property on \textit{injectivity} of the mixing function as one our main goals, but we have not achieved that yet, we still have to use the technique of \textit{model counting} in order to conclude the same. However, the idea of \textit{Formula-2} (\cref{eq:2}) being \textit{satisfiable} is an interesting result and this is where we need to dig deeper.                                                                                                                             

We were not able to make too much progress on our second goal which was to find the longest chain of inputs to retrieve back the original state, and we have to figure out a method to test this out in future.



\clearpage
\renewcommand*{\thesection}{}\textbf{}

\bibliographystyle{alpha}
\bibliography{Bibliography.bib}

\end{document}